\documentclass[draftcls,onecolumn,peerview,12pt]{IEEEtran}

\usepackage{cite}      

\usepackage{graphicx}  

\begin{document}
%
\title{Bounds on the Maximum Number of Concurrent Links in MIMO Ad Hoc
Networks with QoS Constraints}
%
%
\author{
        \normalsize{Pengkai Zhao, and Babak Daneshrad\\
        Wireless Integrated Systems Research (WISR) Group, Electrical Engineering Department,\\
        University of California, Los Angeles, CA 90095 USA\\
        (e-mail: pengkai@ee.ucla.edu; babak@ee.ucla.edu)}}

\maketitle

\begin{abstract}
Multiple-Input Multiple-Output (MIMO) based Medium Access Control
(MAC) protocols have received a good deal of attention as
researchers look to enhance overall performance of Ad Hoc networks
by leveraging multi antenna enabled nodes \cite{1_Fair_MAC,
2_nullhoc, 3_SPACEMAC2005, 4_Adaptive_Interference_Cancellation2005,
5_cross_layer}. To date such MAC protocols have been evaluated
through comparative simulation based studies that report on the
number of concurrent links the protocol can support. However, a
bound on the maximum number of concurrent links (MNCL) that a MIMO
based MAC protocol should strive to achieve has hitherto been
unavailable. In this paper we present a theoretical formulation for
calculating the bound on the MNCL in a Mobile Ad Hoc Network (MANET)
where the nodes have multiple antenna capability, while guaranteeing
a minimum Quality of Service (QoS). In an attempt to make our
findings as practical and realistic as possible, the study
incorporates models for the following PHY layer and channel
dependent elements: (a) path loss and fast fading effects, in order
to accurately model adjacent link interference; (b) a Minimum Mean
Squared Error (MMSE) based detector in the receiver which provides a
balance between completely nulling of neighboring interference and
hardware complexity. In calculating the bound on the MNCL our work
also delivers the optimal power control solution for the network as
well as the optimal link selection. The results are readily
applicable to MIMO systems using Receive Diversity, Space Time Block
Coding (STBC), and Transmit Beamforming and show that with a 4
element antenna system, as much as $3\times$ improvement in the
total number of concurrent links can be achieved relative to a SISO
based network. The results also show diminishing improvement as the
number of antennas is increased beyond 4, and the maximum allowable
transmit power is increased beyond 10 dBm (for the simulated
parameters).
\end{abstract}


%
\IEEEpeerreviewmaketitle

\section{Introduction}
Wireless Ad Hoc networking has emerged as an important aspect of
next generation communication systems. For conventional Single-Input
Single-Output (SISO) system, interference among nodes drastically
limits the number of concurrent (simultaneous) links in Ad Hoc
networks. Multi antenna, multi-input-multi-output (MIMO), based
wireless communications has the ability to spatially null
interference and in so doing increase the number of concurrent links
within a Mobile Ad Hoc Networks (MANET), thus increasing overall
network throughput. In fact some work found in the literature
\cite{1_Fair_MAC, 2_nullhoc, 3_SPACEMAC2005,
4_Adaptive_Interference_Cancellation2005, 5_cross_layer,
6_CrossIssue06} look to MIMO capable MAC protocols as a means of
increasing the network efficiency and its sum-throughput.

The maximum number of concurrent links is a metric used in the
literature \cite{1_Fair_MAC, 8_Mesh, 9_interference_temp} to
evaluate the capacity of a network. Examples of MACs that support
concurrent links in a network where all nodes have multiple antennas
at their disposal can be found in \cite{1_Fair_MAC, 2_nullhoc,
3_SPACEMAC2005, 4_Adaptive_Interference_Cancellation2005,
5_cross_layer, 8_Mesh}. For convenience these MAC protocols will be
referred to as concurrent-based MACs in this paper. The Null-Hoc
\cite{2_nullhoc} and SPACE-MAC \cite{3_SPACEMAC2005} protocols look
to enable concurrent links by using the Gram-Schmidt
Orthonormalization, so as to create orthogonal channels among
different links. The protocol in
\cite{4_Adaptive_Interference_Cancellation2005} uses adaptive
interference cancellation both at the transmitter and at the
receiver, as well as a simple power control protocol for each link.
Multiple links are assumed to access the channel sequentially and
work simultaneously. The MIMA-MAC protocol \cite{5_cross_layer} uses
space division multiple access techniques to guarantee the
concurrency of different communicating links in the network.

Although these concurrent-based MAC protocols have proved to
outperform the conventional SISO based MACs such as the IEEE 802.11
DCF \cite{7_MAC}, a natural question to ask is that how close they
actually come to the theoretical bound (limit) of concurrency in Ad
Hoc networks. Furthermore, since MIMO systems enable a variety of
approaches in utilizing multiple antennas in the physical layer
\cite{19_MIMO}, it is also concerned that how this bound of
concurrency is affected by the choice of MIMO algorithms and
associated physical layer techniques. In this work we identify the
theoretical Maximum Number of Concurrent Links (MNCL) in the network
by considering the following PHY layer and channel dependent
elements: (a) path loss and fast fading effects; (b) different MIMO
transmit/receive algorithms; (c) a Minimum Mean Squared Error (MMSE)
based detector in the receiver; (d) optimal power control and
optimal link selection. The derived MNCL acts as a performance
benchmark for concurrent-based MAC protocols, and is also used as a
metric for comparing different MIMO techniques and parameters.


Our study is based on the assumption that each transmit/receive pair
requires the same Constant Bit Rate (CBR). In this way, the MNCL
that can be had in a MIMO capable network is identified subject to a
minimum Quality of Service (QoS) constraint. In our case the QoS
constraint is the received Signal to Interference plus Noise Ratio
(SINR) which is directly related to Bit Error Rate (BER) and Packet
Error Rate (PER). The proposed framework is executed via an
iterative process for power allocation and a Backtracking-based
search strategy for link selection. Results of different MIMO
transmit algorithms such as Receive Diversity, Space-Time Block
Coding (STBC) and Transmit Beamforming on the MNCL bound are
studied. Using this bound we present design-relevant insight
regarding the impact of the number of pairs, the number of antennas,
and the maximum allowable transmit power per pair on the MNCL.

The reminder of this paper is organized as follows. The proposed
system model is described in Section II. The definition of
concurrent links is provided in Section III. Optimal power
allocation is investigated in Section IV. In Section V, we propose a
backtracking-based strategy to find the optimal link selection.
Numerical results are shown in Section VI, and we conclude in
Section VII.

\section{System Model}
\subsection{Network Model}
We focus on a small network (or sub-network), where each node is
within the transmission range of any other node subject to path loss
and Rayleigh flat fading (Fig. 1). Assume $K$ simultaneously
communicating pairs in the network. Each pair is composed of one
transmit node and one receive node, which are all randomly
distributed in the network. Nodes in different pairs are unique and
independent. They are all equipped with $M$ receive antennas,
sharing the same frequency band and requiring the same Constant Bit
Rate (CBR). The number of transmit antennas varies depending on the
MIMO technique being considered. For simplicity, each communicating
pair is labeled as a \textit{Transceiver Pair}. For the $k$th
transceiver pair, the transmit node and receive node are named
\textit{Tx Node $k$ } and \textit{Rx Node $k$}, respectively.

The propagation between nodes is characterized by path loss and
Rayleigh fading \cite{10_rate_regions, 11_spatial_reuse,
12_Wireless_Rappaport}. For path loss, we use the simplified model
in \cite{12_Wireless_Rappaport}, which is:

\begin{equation}
L_{P}(d) (\textrm{dB}) = L_{P}(d_{0})+10\alpha\log_{10} { d \over
d_{0} }
\end{equation}

In this model, $\alpha$ is the path loss exponent, $d_{0}$ is the
reference distance, and $d$ is the distance between nodes obtained
from node topology information. Both $\alpha $ and $d_{0}$ are
parameters in our study and can be set by the user. We use
$\rho_{kj}$ to denote the power loss ratio from Tx Node $j$ to Rx
Node $k$, and $\rho_{kj}=10^{-\frac{L_{P}(d)}{10}}$.

We assume a flat fading environment, which is modeled by a Rayleigh
distribution. $\textbf{H}_{kj}(m)$ denotes the Rayleigh fading
channel from the $m$th antenna of Tx Node $j$ to Rx Node $k$. It is
an $M\times 1$ vector and consists of independent identically
distributed (i.i.d.) complex Gaussian random variables.

The background white noise is a circularly complex Gaussian vector
with covariance matrix $\sigma_{N}^{2}\textbf{I}_{M}$.
$\textbf{I}_{M}$ is an $M \times M$ unitary matrix, and $\sigma_N^2$
is given by:

\begin{equation}
\sigma_{N}^{2}\textrm{(dBm)} = \eta_{n} + 10\log_{10}(W) + F_{n}
\end{equation}
$W$(Hz) is the bandwidth of the system, while $\eta_{n}$(dBm/Hz) and
$F_{n}$(dB) are the power spectral density of the thermal noise and
the noise figure of the receiver, respectively. They are assumed to
be identical for all the nodes.

\subsection{Transceiver Model}
Three different MIMO techniques are considered in this paper, they
are: (a) $1 \times M$ Receive Diversity; (b) $2 \times M$ Space-Time
Block Coding (STBC); and (c) $M \times M$ Transmit Beamforming. In
this paper, we assume that data symbols in the baseband have the
same modulation type, regardless of single or multiple antennas. For
fair comparison, the number of transmit antennas in these MIMO
techniques are selected so that their spectral efficiency is the
same as that of a Single-Input Single-Output (SISO) system.

\subsubsection{$1 \times M$ Receive Diversity}

In the $1 \times M$ scenario \cite{14_Wireless_Comm}, only one
transmit antenna is employed. Without loss of generality, the $1$st
transmit antenna is used. Considering the $k$th transceiver pair,
the received signal at Rx Node $k$ is given by:
\begin{equation}
{\rm {\bf Y}}_k =\sqrt {P_k \rho _{kk} } {\rm {\bf H}}_{kk} (1)x_k
+{\rm {\bf R}}_k^{(SD)} +{\rm {\bf N}}_k
\end{equation}
\begin{equation}
{\rm {\bf R}}_k^{(SD)} = \sum_{i=1,i\neq k}^{K} \sqrt {P_i \rho
_{ki} } {\rm {\bf H}}_{ki}(1)x_i
\end{equation}

In these equations, $\rm{\bf Y}$, $\rm{\bf H}$, $\rm{\bf R}$ and
$\rm{\bf N}$ are all $M\times 1$ vectors. $\rm{\bf R}$ denotes the
inter-pair interference. $x_k$ is the transmitted symbols for the
$k$th pair, which has zero mean and unit variance. $P_{k}$ is the
allocated power for $x_k$. ${\rm{\bf N}}_k$ is a white Gaussian
noise vector with covariance $\sigma_{N}^{2}\textbf{I}_{M}$.

\subsubsection{$2 \times M$ STBC}

In the $2 \times M$ Space-Time Block Coding (STBC) scenario we use
an Alamouti Code \cite{17_alamouti} with two transmit antennas. We
use $x_{k,m,n} $ to denote the transmitted symbol at $m$th antenna
and $n$th time slot of the $k$th pair. Using the Alamouti Code, we
have $x_{k,1,2} =-x_{k,2,1}^\ast $ and $x_{k,2,2} =x_{k,1,1}^\ast $,
$(1\le m\le 2,1\le n\le 2)$. Here $x_{k,1,1} $ and $x_{k,2,1} $ are
independent symbols with zero mean and unit variance, and $(\cdot
)^\ast $ is the complex conjugation. $P_k(1)$ and $P_k(2)$ are
allocated power for $x_{k,1,1} $ and $x_{k,2,1} $, respectively.
Assume the received vectors corresponding to these two time slots
are ${\rm{\bf Y}}_{k,1}$ and ${\rm{\bf Y}}_{k,2}$, as well as the
noise vectors ${\rm{\bf N}}_{k,1}$ and ${\rm{\bf N}}_{k,2}$. Define
new vectors by $\overline{\rm{\bf Y}}_k = \left[{\rm{\bf
Y}}_{k,1};{\rm{\bf Y}}_{k,2}^{\ast};\right]$, $\overline{\rm{\bf
H}}_{ki,1} = \left[{\rm{\bf H}}_{ki}(1);{\rm{\bf
H}}_{ki}(2)^{\ast};\right]$, $\overline{\rm{\bf H}}_{ki,2} =
\left[{\rm{\bf H}}_{ki}(2);-{\rm{\bf H}}_{ki}(1)^{\ast};\right]$ and
$\overline{\rm{\bf N}}_k=[{\rm{\bf N}}_{k,1};{\rm{\bf
N}}_{k,2}^{\ast};]$. Consequently, the received signal at Rx Node
$k$ can be denoted as:
\begin{equation}
\overline {\rm {\bf Y}} _k =\sqrt {P_k (1)\rho _{kk} } \overline
{\rm {\bf H}} _{kk,1} x_{k,1,1} +\sqrt {P_k (2)\rho _{kk} }
\overline {\rm {\bf H}} _{kk,2} x_{k,2,1} +{\rm {\bf R}}_k^{(STBC)}
+{\rm {\bf N}}_k
\end{equation}
\begin{equation}
{\rm {\bf R}}_k^{(STBC)} =\sum\limits_{i=1,i\ne k}^K
{\sum\limits_{l=1}^2 {\sqrt {P_i (l)\rho _{ki} } \overline {\rm {\bf
H}} _{ki,l} x_{i,l,1} } }
\end{equation}

\subsubsection{$M \times M$ Transmit Beamforming}

The Beamforming method outlined in \cite{15_Chen2006, 18_coded_MIMO}
is also employed in this paper. $M$ transmit antennas are adopted,
and an $M \times 1$ weight vector ${\rm{\bf u}}_{k}$ is applied to
these transmit antennas. Let ${\rm{\bf H}}_{kk}=\left[ {\rm{\bf
H}}_{kk}(1),{\rm{\bf H}}_{kk}(2),\ldots,{\rm{\bf
H}}_{kk}(M)\right]$, then according to \cite{15_Chen2006,
18_coded_MIMO}, ${\rm{\bf u}}_k$ is the eigenvector corresponding to
the largest eigenvalue of the matrix ${\rm{\bf H}}_{kk}^{H}{\rm{\bf
H}}_{kk}$, where $(\cdot)^{H}$ denotes conjugate transpose. We
assume that ${\rm{\bf u}}_k^{H}{\rm{\bf u}}_k=1$ and $x_k$ has zero
mean and unit variance. The transmit power for $x_k$ is $P_k$. Then
the received signal under this method is represented as:
\begin{equation}
\mbox{Y}_k =\sqrt {P_k \rho _{kk} } {\rm {\bf H}}_{kk} {\rm {\bf
u}}_k x_k +{\rm {\bf R}}_k^{(Beam)} +{\rm {\bf N}}_k
\end{equation}
\begin{equation}
{\rm {\bf R}}_k^{(Beam)} =\sum\limits_{i=1,i\ne k}^K {\sqrt {P_i
\rho _{ki} } {\rm {\bf H}}_{ki} {\rm {\bf u}}_i x_i }
\end{equation}

\subsection{MMSE Solution}
For all three scenarios listed in the previous section we will use
the Minimum Mean Squared Error (MMSE) solution at the receiver
\cite{13_JointOptimal1998, 20_MMSE, 21_Spatial_Multiplexing} to
arrive at the receive MIMO antenna weights. Here we only present the
MMSE solution for the $1 \times M$ Receive Diversity case. However,
the MMSE solution can also be derived for the $2 \times M$ STBC and
$M \times M$ Transmit Beamforming scenarios by slight modification
(the associated results are given in Appendix I).

For the $1 \times M$ Receive Diversity scenario, the estimate of the
transmitted stream $x_{k}$ at the receiver is given by (\ref{eqn_9})
where the vector ${\rm {\bf w}}_k $ is the MMSE solution.

\begin{equation}
\label{eqn_9} \hat{x}_k = {\rm{\bf w}}_k^{H} {\rm{\bf Y}}_k
\end{equation}

The corresponding SINR at the receiver is:
\begin{equation}
\Gamma _k =\frac{P_k \rho _{kk} {\rm {\bf w}}_k^H {\rm {\bf H}}_{kk}
(1){\rm {\bf H}}_{kk}^H (1){\rm {\bf w}}_k }{{\rm {\bf w}}_k^H
\Phi_{(SD)}(k){\rm {\bf w}}_k }
\end{equation}
\begin{equation}
\label{eqn_12} \Phi_{(SD)}(k) = \sum\limits_{1\leq i \leq K, i \neq
k} P_i \rho_{ki} {\rm{\bf H}}_{ki}(1) {{\rm{\bf H}}_{ki}^{H}}(1) +
\sigma_N^{2}\rm{\bf I}_M
\end{equation}

With the constraint that ${\rm{\bf w}}_k^H{\rm{\bf H}}_{kk}(1)=1$,
optimal linear vector ${\rm{\bf w}}_k$ in the MMSE solution is given
in \cite{13_JointOptimal1998, 20_MMSE, 21_Spatial_Multiplexing}

\begin{equation}
\label{eqn_11} {\rm{\bf w}}_k=\frac{\Phi_{(SD)}^{-1}(k){\rm{\bf
H}}_{kk}(1)}{{\rm{\bf H}}_{kk}^{H}(1)\Phi_{(SD)}^{-1}(k){\rm{\bf
H}}_{kk}(1)}
\end{equation}

The SINR with MMSE solution $\hat{\Gamma}_k$ is:
\begin{equation}
\hat{\Gamma}_k = P_k \rho_{kk}{\rm{\bf H}}_{kk}^{H}(1)
\Phi_{(SD)}^{-1}(k){\rm{\bf H}}_{kk}(1)
\end{equation}

Finally, we assume a QoS requirement at the receiver. Consider an
SINR threshold $\gamma_T$, for the $k$th transceiver pair, they can
be correctly received if and only if the received SINR is not lower
than $\gamma_T$. The threshold $\gamma_T$ represents the QoS
requirement and is being used here in the same manner as the
Physical Model in \cite{16_Gupta2000}.

\section{The Definition of Concurrent Links}
We assume that all Tx and Rx Nodes are mobile, and their locations
are randomly distributed and varying with time. In the following
discussion, a specific scenario means one specific realization of
node locations in the region, and the channel responses between
them. For a specific scenario, given the allocated power in each
pair, the SINR for each pair can be evaluated using the results of
Section II. Based on these SINR results, we first provide the
following definition.
\newtheorem{definition}{Definition}
\begin{definition}[Link]
Consider a specific scenario with $K$ transceiver pairs. The
transmit power for every pair is constrained by $P_T$. The $k$th
pair $(1\leq k \leq K)$ is called a \textit{link} if and only if it
satisfies the following conditions:

For $1 \times M$ Receive Diversity and $M \times M$ Transmit
Beamforming:

\begin{equation}
\hat{\Gamma}_k\geq \gamma_T \textrm{ and } P_k \leq P_T
\end{equation}

For Space Time Block Coding:

\begin{equation}
\hat{\Gamma}_k(l)\geq \gamma_T, 1 \leq l \leq 2 \textrm{ and }
\sum\limits_{l=1}^{2}P_k(l)\leq P_T
\end{equation}
\end{definition}

Assume a specific scenario with $K$ transceiver pairs, labeled from
$1$ to $K$ and denoted by a set $U=\{1,2,\ldots,K\}$. A pair set
$U_P \subseteq U$ is a feasible pair set if there exists a power
allocation for all transmitters in $U_P$ such that
\textit{Definition 1} holds for all transceiver pairs in $U_P$
(i.e., the QoS constraint is satisfied at all receivers subject to
the maximum $P_T$ constraint). Then all the pairs in $U_P$ are named
\textit{Concurrent Links}.

Let us denote the number of pairs in $U_P$ as: $|U_P|$. Then the
MNCL is calculated by the following optimization problem:

\begin{eqnarray}
\label{eqn_16} \max& |U_P| \nonumber\\
\textrm{s.t.}& U_P\subseteq U \textrm{ and } U_P \textrm{ is
feasible}
\end{eqnarray}

For notational simplicity, the result of the above optimization
problem is denoted as $N_{\max}(K,M)$. Note that several different
feasible sets $U_P$ may produce the same $N_{\max}(K,M)$. We then
define \textit{Average MNCL} as the expectation of $N_{\max}(K,M)$,
averaged over random locations and random channel responses. The
Average MNCL will be denoted as $C(K,M)$ in this paper.

The optimization in equation (\ref{eqn_16}) can be divided into two
steps: 1) to examine whether a pair vector $U_P$ is feasible; and 2)
to search for the feasible pair set $U_P$ with the MNCL. In this
paper, the first step is solved by an optimal power allocation
process presented in Section IV, while the second step, referred to
as optimal link selection, is solved in Section V.

\section{Optimal Power Allocation}
Optimal power allocation is an important factor when deriving the
bound on the MNCL. In our formulation it is used to decide whether a
given pair set $U_P$ is feasible per \textit{Definition 1}. In this
section, the algorithm for optimal power allocation is presented
from conventional power control techniques
\cite{13_JointOptimal1998, 22_Powercontrol1995,
23_FrameworkUplink1995}. Without loss of generality, only the $1
\times M$ Receive Diversity scenario is considered in this section.
However, the proposed algorithm can also be applied to the $2 \times
M$ STBC and $M \times M$ Transmit Beamforming scenarios after slight
modification.

\subsection{Iterative Power Control}
Consider $K$ transceiver pairs in the network using $1 \times M$
Receive Diversity. The allocated power for each pair is stacked into
a vector $\rm{\bf P}$, which is defined as the power vector and
given by:

\begin{equation}
{\rm{\bf P}} = \left[ P_1,P_2,\ldots,P_K\right]
\end{equation}
Since we only focus on pairs in $U_P$, we have the following
constraint:
\begin{equation}
\label{Pk0} \mbox{for }k\notin U_P ,\mbox{ }P_k =0,\mbox{ }1\le k\le
K
\end{equation}

Let ${\rm{\bf w}}_k$ be the MMSE solution in equation
(\ref{eqn_11}). We define the following iteration equation:

For $k \in U_P$

\begin{equation}
P_{k}^{n+1}=\gamma_T \frac{C_k^{(SD)}\left\{{\rm{\bf w}}_k,{\rm{\bf
P}}^{n}\right\}+\sigma_N^{2}{\rm{\bf w}}_k^{H}{\rm{\bf
w}}_k}{\rho_{kk}}
\end{equation}

\begin{equation}
C_k^{(SD)}\left\{{\rm{\bf w}}_k, {\rm{\bf
P}}^{n}\right\}=\sum\limits_{i=1,i\neq
k}^{K}P_i^{n}\rho_{ki}G\left\{{\rm{\bf w}}_k, {\rm{\bf
H}}_{ki}(1)\right\}
\end{equation}
\begin{equation}
G\left\{ {{\rm {\bf w}}_k ,{\rm {\bf H}}_{ki} (1)} \right\}={\rm
{\bf w}}_k^H {\rm {\bf H}}_{ki} (1){\rm {\bf H}}_{ki}^H (1){\rm {\bf
w}}_k
\end{equation}
For $1\leq k \leq K$ and $k \notin U_P$

\begin{equation}
P_k^{n+1}=0
\end{equation}
where $n$ denotes the $n$th iteration. For simplicity, the above
iteration is denoted as

\begin{equation}
\label{eqn_23} {\rm{\bf P}}^{n+1}=m\left({\rm{\bf P}}^{n}\right)
\end{equation}
Here ${\rm{\bf P}}^{n}$ is the power vector for the $n$th iteration.

We define the fixed point of mapping as the power vector
$\widehat{\rm{\bf P}}$ satisfying $\widehat{{\rm{\bf
P}}}=m(\widehat{{\rm{\bf P}}})$. The following theorem holds for the
iterative equation (\ref{eqn_23}), which is used to verify the
existence of optimal power allocation in this paper.

\newtheorem{theorem}{Theorem}
\begin{theorem}
Given $K$ transceiver pairs and a specific pair set $U_P$. If $U_P$
is feasible, then a unique fixed point of mapping, $\widehat{\rm{\bf
P}}$, exists that satisfies $\widehat{\rm{\bf P}}=m(\widehat{\rm{\bf
P}})$ and $\widehat{P}_k\leq P_T, \forall k \in U_P$. Furthermore,
corresponding to the unique power vector $\widehat{\rm{\bf P}}$ is a
unique receive weight vector $\widehat{\rm{\bf w}}_k$ given by the
MMSE solution.
\end{theorem}

\begin{proof}
Please refer to Appendix II. Note that the converse proposition of
this theorem holds obviously. That is, if there exists
$\widehat{\rm{\bf P}}=m(\widehat{\rm{\bf P}})$ and
$\widehat{P}_k\leq P_T, \forall k \in U_P$, then $U_P$ is feasible.
\end{proof}



\subsection{Decision Criteria}
Theorem 1 suggests that a feasible pair set can be identified if and
only if a fixed point of mapping for the transmit power levels
exists and the power constraints are met. Moreover,
\cite{13_JointOptimal1998} suggests that the power control algorithm
in (\ref{eqn_23}) is guaranteed to converge to a fixed point of
mapping even when starting from an arbitrary initial power vector
${\rm{\bf P}}^0$. This suggests a rather straight forward approach
for determining the feasibility of a given pair set $U_P$ as
captured by the following three criteria.

\newtheorem{criterion}{Criterion}
\begin{criterion}
Assume that the iteration process starts from initial condition
${\rm{\bf P}}^0 = 0$. In each iteration, if the power in any
transceiver pair exceeds the power constraint $P_T$, then $U_P$ is
not feasible.
\end{criterion}

The proof of Criterion 1 is referred to Theorem 2 in
\cite{13_JointOptimal1998}.

\begin{criterion}
Assume that the iteration process starts from arbitrary condition
${\rm{\bf P}}^0$. In each iteration, if ${\rm{\bf P}}^n$ is
feasible, then these pairs can be supported simultaneously.
\end{criterion}

Here, saying ${\rm{\bf P}}^n$ is feasible means that by using
${\rm{\bf P}}^n$, for $1 \leq k \leq K$, we can have
$\hat{\Gamma}_k\geq \gamma_T$ and $P_k^n \leq P_T$. This criterion
is straightforward and it is helpful in reducing the iterations in
the decision process.

\begin{criterion}
After $I_{\max}$ number of iterations, if the conditions in
\textit{Criterion 2} have not yet been satisfied for pairs in $U_P$,
then we declare that $U_P$ is not feasible.
\end{criterion}

Criterion 3 guarantees that the decision can be made within a finite
iteration number $I_{\max}$. Using this criterion, some feasible
pair sets may be missed. However, proper selection of $I_{\max}$ can
drive the probability of missing a feasible pair set to be
arbitrarily close to zero.

Using the above criteria we now outline a sequential procedure by
which we can automatically determine if a set $U_{P}$ is feasible.
The steps are as follows:

\textit{Iterative Determination of Feasibility (IDF)}
\begin{enumerate}
\item Given pair set $U_P$. Initialize $n=0$ and ${\rm{\bf P}}^0 =
0$;

\item Iterate by ${\rm{\bf P}}^{n+1}=m\left({\rm{\bf
P}}^n\right)$. Using the updated power vector, ${\rm{\bf P}}^{n+1}$,
calculate the SINR for each pair under the MMSE solution;

\item If $U_P$ is not feasible by Criterion 1, then go to step 6;
else, go to step 4;

\item If $U_P$ is feasible by Criterion 2, then go to step 6; else,
go to step 5;

\item If $U_P$ is not feasible by Criterion 3, then go to step 6;
else, $n=n+1$, go to step 2.

\item The feasibility of $U_P$ is determined, stop.
\end{enumerate}

\section{Finding $U_{P}$ with Maximum Number of Concurrent Links}
In the previous section, we derived the method to determine if a
pair set is feasible. In this section we describe how to find the
feasible pair set with the MNCL. Naturally, a brute force search can
be employed for this problem. However, in order to reduce the search
space, we first characterize the property of the feasible pair set.

Consider a total of $K$ transceiver pairs, labeled $1$ to $K$.
Define $U=\left\{1,2,\ldots,K\right\}$ and $U_P$ is a given pair
set.

\begin{theorem}
For two pair sets $U_{P,2}\subseteq U_{P,1}$, if $U_{P,1}$ is
feasible, then $U_{P,2}$ is also feasible.
\end{theorem}

\begin{proof}
Consider the $1\times M$ Receive Diversity scenario. Let
$\left(U_{P,1}-U_{P,2}\right)$ denote the pairs in $U_{P,1}$ but not
in $U_{P,2}$. If $U_{P,1}$ is feasible with the associated power
vector $\rm{\bf P}$, then for the pairs in $U_{P,2}$, keep the power
$P_k$ and linear vector ${\rm{\bf w}}_k$ unchanged. Meanwhile, shut
down the pairs in $\left(U_{P,1}-U_{P,2}\right)$. Here,
\textit{shutting down} means setting the corresponding transmit
power to zero. As a result, there are less interference in
${\Phi_{(SD)}(k)}$ of equation (10), and conditions in
\textit{Definition 1} are satisfied for the pairs in $U_{P,2}$. Thus
$U_{P,2}$ is feasible and the above theorem holds. The proof is
extendable for STBC and Beamforming methods as well.
\end{proof}

Theorem 2 shows that the backtracking formulation in
\cite{9_interference_temp, 24_computer_algorithms} can be adopted to
solve the problem of finding the $U_P$ with the MNCL. We start this
formulation with an empty pair set. At every level, each feasible
subset is expanded by including one more pair, constructing new
subsets to be validated by the IDF iterations presented in Section
IV (namely, forward search). If one subset becomes unfeasible, the
algorithm backtracks by removing the trailing pair from the subset,
and then proceeds by expanding the subset with alternative pairs
(namely, backward search). Specifically, we use a depth-first search
strategy to execute this procedure. An example with 4 pairs are
illustrated in Fig. 2, where all feasible pair sets are outlined.
Here the initial feasible subsets are $\{ \{1\}, \{2\}, \{3\}, \{4\}
\}$, while the final feasible subsets are $\{ \{1,2,3\}, \{1,3\},
\{2,3\}, \{2,4\}, \{3,4\}, \{4\} \}$, and the MNCL is 3.

Let $U_{P,search}$ denote the pair set candidate during the search,
and $\left| {U_{P,search} } \right|$ is the number of elements in
$U_{P,search} $. Meanwhile, use $direction=1$ to represent the
forward search, and $direction=0$ the backward search. The
pseudocode of the proposed algorithm is described in the following.

\textit{Backtracking-based Optimal Link Selection (BOLS)}

\begin{enumerate}
\item Given $K$ pairs (labeled from 1 to $K$) in the network and $M$ receive antennas per pair. Initialize by $U_{P,search} =\{1\}$, $N_{\max}(K,M)=0$.
\item If $\left| {U_{p,search} } \right|>N_{\max } (K,M)$, then:
    \begin{enumerate}
    \item Use the IDF process in Section IV to examine whether $U_{P,search} $ is feasible. If $U_{P,search} $ is feasible, then set $N_{\max } (K,M)=\left| {U_{P,search} } \right|$ and $direction=1$; otherwise, set $direction=0$.
    \end{enumerate}
\item If $\left| {U_{p,search} } \right|\le N_{\max } (K,M)$, then set $direction=1$.
\item Update the pair set candidate $U_{P,search} $:
    \begin{enumerate}
    \item If $U_{P,search} =\{K\}$, then all the search have been done, return $N_{\max}(K,M)$ and stop.
    \item If $U_{P,search} \ne \{K\}$, then $U_{P,search}={\rm{\bf PairSet\_Gen}}\left[ {U_{P,search}, direction} \right]$ and go to
step 2.
    \end{enumerate}
\end{enumerate}

In step 3, we only examine the pair set $U_{P,search} $ in which
$\left| {U_{P,search} } \right|$ is larger than the current value of
$N_{\max}(K,M)$. In other words, if $\left| {U_{P,search} }
\right|\le N_{\max } (K,M)$, instead of employing the IDF
procedure$,$ we set $direction=1$ and go to step 4 directly. Next,
the function PairSet{\_}Gen is described as follows:

\textit{Function }${\rm{\bf PairSet\_Gen}}$ 

\begin{enumerate}
\item Input $U_{P,search} $ and $direction$. Find $k_{\max } =\mathop {\max }\limits_{k\in U_{P,search} } k$, which represnts the pair with the maximum index in $U_{P,search} $.
\item If $k_{\max}<K$, then:
    \begin{enumerate}
    \item If $direction=1$, add new element in $U_{P,search} $ by
    $U_{P,search} =U_{P,search} \cup \{k_{\max } +1\}$. Return
    $U_{P,search} $ and stop.

    \item If $direction=0$, update the element $k_{\max}$ in $U_{P,search}$ by $k_{\max}=k_{\max}+1$.
    Return $U_{P,search}$ and stop.
    \end{enumerate}
\item If $k_{\max } =K$ and $\left| {U_{P,search} } \right|>1$, then;
    \begin{enumerate}
    \item Delete $k_{\max}$ from $U_{P,search}$ by $U_{P,search}^{del} =U_{P,search}
-\{k_{\max } \}$. Find $\overline k _{\max } =\mathop {\max
}\limits_{k\in U_{P,search}^{del} } k$, update the element
$\bar{k}_{\max}$ in $U_{P,search}^{del} $ by
$\bar{k}_{\max}=\bar{k}_{\max}+1$.

    \item Set $U_{P,search} =U_{P,search}^{del} $. Return $U_{P,search} $ and
stop.
    \end{enumerate}

\item If $k_{\max } =K$ and $\left| {U_{P,search} } \right|=1$, then
all the search have been done. Return $U_{P,search} $ and stop.
\end{enumerate}

Finally, averaging $N_{\max}(K,M)$ among random locations and random
channel responses, we obtain the $C(K,M)$.

\section{Simulation Results}
The simulation setup randomly distributes the nodes, in accordance
to a uniform distribution, within a disk of radius 100 meters.
Numerical results are averaged over 1000 Monte Carlo simulations of
independent realizations of node topology and channel response. We
assume QPSK modulation in the baseband, and the desired SINR
threshold, $\gamma_{T}$, is 10dB. This corresponds to an uncoded BER
of less than 1e-3 (see Table 6.1 in \cite{14_Wireless_Comm}). The
remaining parameters are listed in Table I.

\subsection{Comparison of MIMO Techniques}
In this section we evaluate the performance of the proposed
algorithms, and use the MNCL as a metric to compare different MIMO
methods. The three MIMO methods outlined in Section II are compared,
namely, $1\times M$ Receive Diversity, $2 \times M$ STBC, and $M
\times M$ Transmit Beamforming, all with MMSE decoding.

We assume up to 15 transceiver pairs, and vary the number of receive
antennas from 1 to 4. Results for the three different MIMO methods
are reported in Fig. 3 and Fig. 4. We observe that Transmit
Beamforming has the best performance due to the fact that it
explores Channel State Information (CSI) at the transmitter. On the
other hand, STBC has a worse performance than Receive Diversity.
Note that STBC is conventionally designed to cope with the white
Gaussian noise \cite{17_alamouti}, but in this study we use MIMO
system to combat the colored interference signals (equation
(\ref{eqn_12})), and our results indicate that Receive Diversity
outperforms STBC in this case. Finally, performance of conventional
SISO system corresponds to that of Receive Diversity with $M=1$. We
observe that compared to a SISO system, as much as $3\times$
improvement in MNCL can be achieved by using $M=4$ receive antennas
(compare the highlighted values for $K=5, 10, 15$ in Fig. 3 with
those in Fig. 4).

Next, we study the convergence of the estimate for the MNCL. We
assume a total of 12 transceiver pairs in the network, with the
number of receive antennas fixed 4. As previously mentioned, we
create a realization by distributing the pairs in a 100m radius disk
and randomly generating the channel responses among them. In Fig. 5,
we plot the average number of MNCL versus the number of
realizations. We find that the simulation results become convergent
after around 500 realizations. Note that we used 1000 independent
realizations in these simulations, which is large enough to yield a
precise estimate for the performance of the proposed algorithm.

Finally, we compare the proposed backtracking strategy (Section V)
with the brute force search. We have verified that, for $1\leq K
\leq 15$, the proposed backtracking method has exactly the same MNCL
as the brute force search (the results are omitted due to space
limit). Here we focus on the search complexity of these two search
methods, that is, the number of times the IDF iterations were
called. The average result over 1000 Monte Carlo simulations is
shown in Fig. 6. Note that the associated complexity for the brute
force method is $\mbox{2}^K$, where $K$ is the number of pairs in
the network. Compared with the brute force search, we observe a
significant reduction in complexity when the backtracking scheme is
used. This verifies the efficacy of our proposed scheme.

\subsection{Impact of Simulation Parameters}
Using the formulation developed in this work, we can investigate the
impact of different parameters, such as the number of transceiver
pairs, the number of receive antennas, and the maximum allowable
transmit power on the MNCLs. We first look at the number of pairs in
the network. Results in Fig. 3 and Fig. 4 have already shown that
the MNCL increases substantially with the number of pairs. Actually,
we can use a two-stage linear equation to approximate the result of
the MNCL. The equation is presented in (\ref{eq8}), and the
approximation result for Transmit Beamforming is demonstrated in
Fig. 7 and Table II.

For $\mbox{1}\le K\le 15$ ($a_1 ,b_1 ,a_2 ,b_2 $ are parameters to
be fitted):
\begin{equation}
\label{eq8} C(K,M)=\left\{ {{\begin{array}{*{20}c}
 {a_1 +b_1 K,} \hfill & {\mbox{if }1\le K\le M} \hfill \\
 {a_2 +b_2 K,} \hfill & {\mbox{if }M+1\le K\le 15} \hfill \\
\end{array} }} \right.
\end{equation}
In the above equation, when $1\le K\le M$, the improvement is mainly
from the diversity gain; while when $M+1\le K\le 15$, the
improvement is from multi-user diversity (as the number of pairs
wanting to communicate increases the likelihood of choosing a subset
of these pairs that exhibit good interference properties increases).
The results in Fig. 3 and Fig. 4 show that the MNCL is larger than
$M$ when the number of pairs $K$ is increased beyond $M$. This
observation highlights some extra gains in MNCL to be exploited by
the MAC protocol.

Next, we focus on the impact of the number of receive antennas. We
assume 12 transceiver pairs, and the number of receive antennas $M$
varies from 1 to 8. Simulation results are shown in Fig. 8. We see
that in this simulation, the MNCL is improved dramatically with the
number of receive antennas. Again, we find that with 4 receive
antennas, the improvement is around 3 times relative to the SISO
system. Furthermore, we observe that saturation starts to set in
when the number of receive antennas is greater than 5. With 8
receive antennas the gain relative to a SISO system is $4\times$.

Lastly, we explore the impact of the maximum allowable transmit
power $P_T$. Naturally, higher transmit power is useful to combat
path loss and increase transmission range. However, here we analyze
the impact of $P_T$ under the assumption that all the nodes are
distributed in a fixed disk with radius 100m. This assumption is
reasonable which corresponds to a geographically constrained area
with many potential transmission pairs (e.g., classroom, conference
room). We assume a total of 10 pairs in the network, with 4 receive
antennas per pair, and consider different power constraints from
-20dBm to 50dBm. The results are depicted in Fig. 9. We observe that
for each MIMO method, initially, the MNCL increases with higher
power constraint. However, when $P_T$ is beyond 10dBm, the
improvement becomes diminishing. The reason is that co-channel
interference among pairs has dominated the network, and increasing
$P_T$ can not mitigate these interfering signals.

\section{CONCLUSION}
In this paper, we have identified the bound on the maximum number of
concurrent links (MNCL) in MIMO Ad Hoc networks subject to a minimum
QoS (SINR in our case) for each link. This number is derived by
considering the practical factors of wireless channel, transmit
power allocation, link selection and MIMO transceiver algorithms in
a realistic MIMO system. We employ an iterative algorithm to examine
the existence of an optimal power allocation that guarantees QoS for
all pairs in a given set. Based on this iterative algorithm, we
proposed a backtracking search algorithm to select the optimal
subset of pairs, constituting the MNCL.

Extensive simulations were conducted to verify the efficacy of the
proposed algorithms and evaluate the impact from different
parameters. The results show a $3\times$ improvement in MNCL with a
4 element antenna system relative to a SISO system. For the
parameters simulated, diminishing improvement is observed when the
number of antennas is increased beyond 5, and when the maximum
transmit power is increased beyond 10dBm.

\appendices
\section{}
In this appendix, we show the MMSE solution for STBC and Transmit
Beamforming methods. For STBC, define that the MMSE solutions for
$x_{k,1,1}$ and $x_{k,2,1}$ are ${\rm{\bf w}}_{k,1}$ and ${\rm{\bf
w}}_{k,2}$, respectively. Then we can have:

\begin{equation}
{\rm{\bf w}}_{k,1}=\frac{\Phi_{(STBC,1)}^{-1}(k)\overline{\rm{\bf
H}}_{kk,1}}{\overline{\rm{\bf
H}}_{kk,1}^H\Phi_{(STBC,1)}^{-1}(k)\overline{\rm{\bf H}}_{kk,1}}
\end{equation}

\begin{equation}
{\rm{\bf w}}_{k,2}=\frac{\Phi_{(STBC,2)}^{-1}(k)\overline{\rm{\bf
H}}_{kk,2}}{\overline{\rm{\bf
H}}_{kk,2}^H\Phi_{(STBC,2)}^{-1}(k)\overline{\rm{\bf H}}_{kk,2}}
\end{equation}

\begin{equation}
\Phi_{(STBC,1)}(k)=\sum\limits_{i=1, i\neq
k}^{K}\sum\limits_{l=1}^{2}P_i(l)\rho_{ki}\overline{\rm{\bf
H}}_{ki,l}\overline{\rm{\bf
H}}_{ki,l}^H+P_k(2)\rho_{kk}\overline{\rm{\bf
H}}_{kk,2}\overline{\rm{\bf H}}_{kk,2}^H+\sigma_N^2\rm{\bf I}_M
\end{equation}

\begin{equation}
\Phi_{(STBC,2)}(k)=\sum\limits_{i=1, i\neq
k}^{K}\sum\limits_{l=1}^{2}P_i(l)\rho_{ki}\overline{\rm{\bf
H}}_{ki,l}\overline{\rm{\bf
H}}_{ki,l}^H+P_k(1)\rho_{kk}\overline{\rm{\bf
H}}_{kk,1}\overline{\rm{\bf H}}_{kk,1}^H+\sigma_N^2\rm{\bf I}_M
\end{equation}

For Transmit Beamforming, the MMSE solution is:
\begin{equation}
{\rm{\bf w}}_{k}=\frac{\Phi _{(BF)}^{-1}(k) {\rm {\bf H}}_{kk} {\rm
{\bf u}}_k}{({\rm {\bf H}}_{kk} {\rm {\bf u}}_k )^H\Phi
_{(BF)}^{-1}(k) {\rm {\bf H}}_{kk} {\rm {\bf u}}_k}
\end{equation}
\begin{equation}
\Phi_{(BF)}(k)=\sum\limits_{i=1,i\ne k}^K {P_i \rho_{ki} ({\rm {\bf
H}}_{ki} {\rm {\bf u}}_k )}({\rm {\bf H}}_{ki} {\rm {\bf u}}_k )^H
+\sigma_N^2 {\rm {\bf I}}_M
\end{equation}

\section{Proof of Theorem 1}
Note that we have assumed in (\ref{Pk0}) that $\{P_k=0, k\notin
U_P\}$. The iteration equation (\ref{eqn_23}) can be proved to be a
standard function \cite{23_FrameworkUplink1995} for $\{P_k, k\in
U_P\}$ via similar manner in \cite{13_JointOptimal1998}. Now assume
that with power vector ${\rm{\bf P}}$, $U_P$ is feasible. Set the
initial power vector ${\rm{\bf P}}^0={\rm{\bf P}}$, and run the
iteration process by ${\rm{\bf P}}^{n+1}=m({\rm{\bf P}}^n)$.
According to Lemma 1 in \cite{23_FrameworkUplink1995}, ${\rm{\bf
P}}^n$ will convergence to the fixed point of mapping, which is
$\widehat{{\rm{\bf P}}}= m(\widehat{{\rm{\bf P}}})$. With
Monotonicity property in standard function, for $1 \leq k \leq K$,
we have $\widehat{P}_k\leq P_k^0 \leq P_T$. Finally, it has been
proved in \cite{13_JointOptimal1998} that power and weight vectors
corresponding to the fixed point of mapping are all unique. Thus the
theorem follows.

\bibliographystyle{IEEEtran}
\bibliography{IEEEabrv,Paper_1_PK}

\newpage
\begin{table}
\caption{Parameters in the Simulations} \centering \label{table I}
\begin{tabular}{c||c}
\hline \textbf{Parameter} & \textbf{Value} \\
\hline \textrm{Path Loss} & $d_{0}=1$\textrm{m} \\
                          & $L_P(d_0)=46$\textrm{dB} \\
                          & \textrm{Exponent factor} $\alpha=3$ \\
\hline \textrm{Noise Power Spectral Density} & $\eta_n=-174$\textrm{dBm/Hz} \\
\hline \textrm{Noise Figure} & $F_n=4$\textrm{dB} \\
\hline \textrm{Bandwidth} & $W=1$\textrm{MHz} \\
\hline \textrm{Maximum Transmit Power} & $P_T=20$\textrm{dBm} \\
\hline \textrm{SINR threshold} &$\gamma_{T}=10$\textrm{dB} \\
\hline
\end{tabular}
\end{table}

\begin{table}
\caption{Parameters in the Approximation Results of Average MNCL
with Transmit Beamforming} \centering \label{table II}
\begin{tabular}{|c|c|c|c|c|}
\hline $M$ & $1$ & $2$ & $3$ & $4$ \\
\hline
\hline $a_1$ & $0$ & $0.0040$ & $-0.0003$ & $0.0030$ \\
\hline $b_1$ & $0.9680$ & $0.9950$ & $0.9990$ & $0.9956$ \\
\hline $a_2$ & $0.9961$ & $1.9587$ & $2.6045$ & $3.5086$ \\
\hline $b_2$ & $0.1084$ & $0.2121$ & $0.3013$ & $0.3171$ \\
\hline
\end{tabular}
\end{table}

\begin{figure}
\centering
\includegraphics[width=4in]{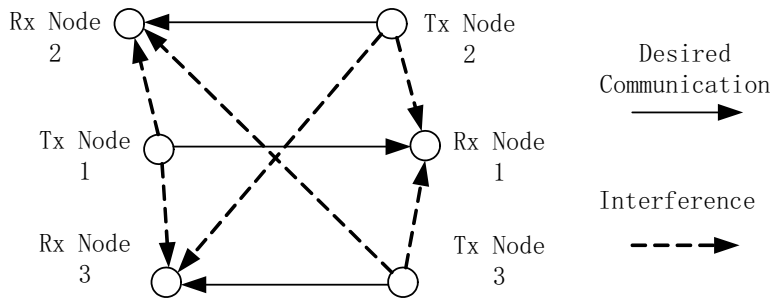}
\caption{Illustration of the considered network. Solid lines are the
desired communication, while dashed lines are the interference.}
\label{Fig1}
\end{figure}

\begin{figure}
\centering
\includegraphics[width=4in]{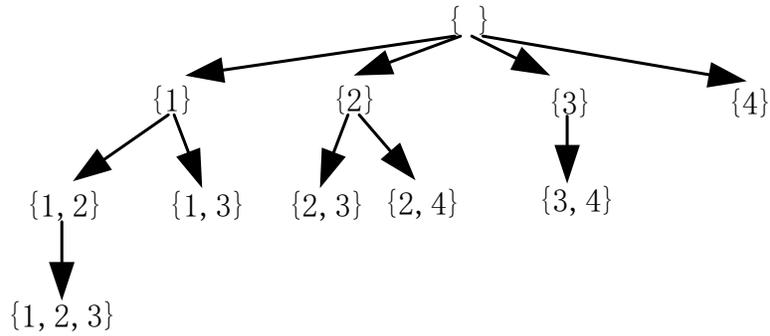}
\caption{Tree structure for backtracking search.} \label{Fig2}
\end{figure}

\begin{figure}
\centering
\includegraphics[width=4in]{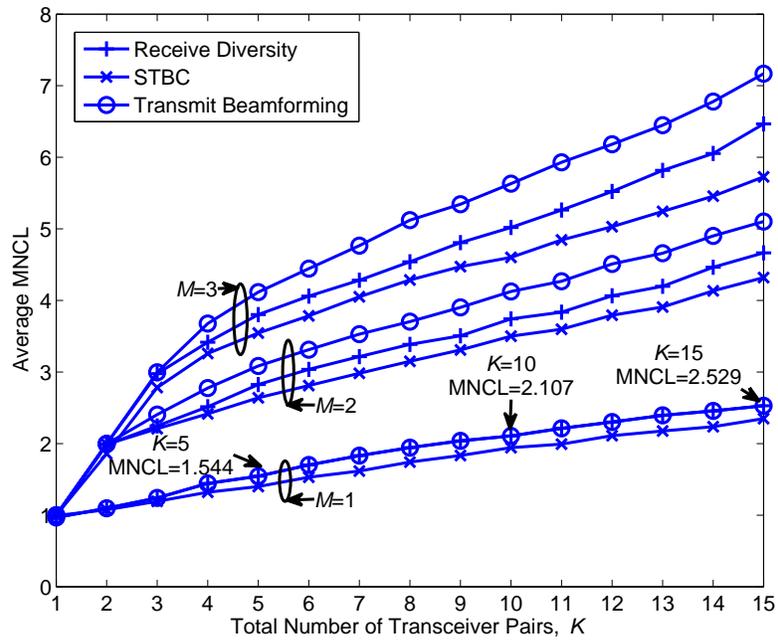}
\caption{Average MNCL under different MIMO techniques. $M$ is from 1
to 3.} \label{Fig3}
\end{figure}

\begin{figure}
\centering
\includegraphics[width=4in]{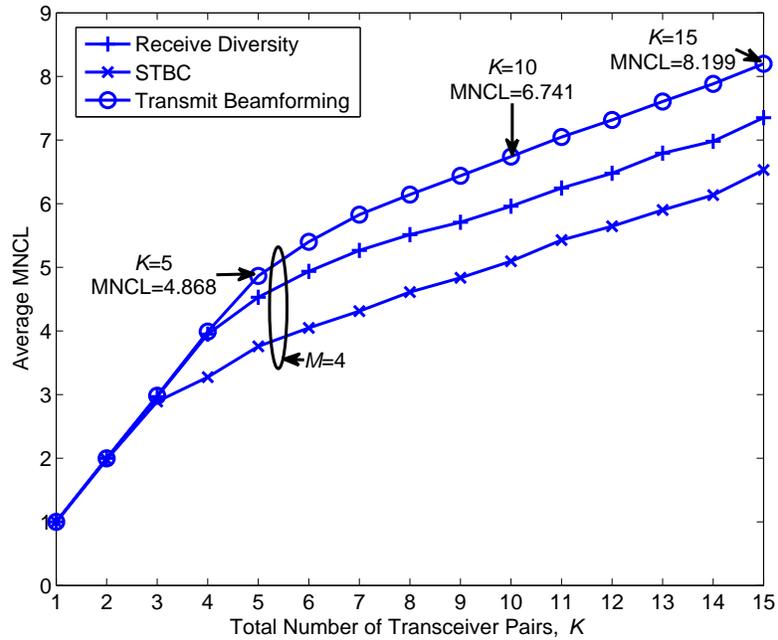}
\caption{Average MNCL under different MIMO techniques. $M$ is set as
4.} \label{Fig4}
\end{figure}

\begin{figure}
\centering
\includegraphics[width=4in]{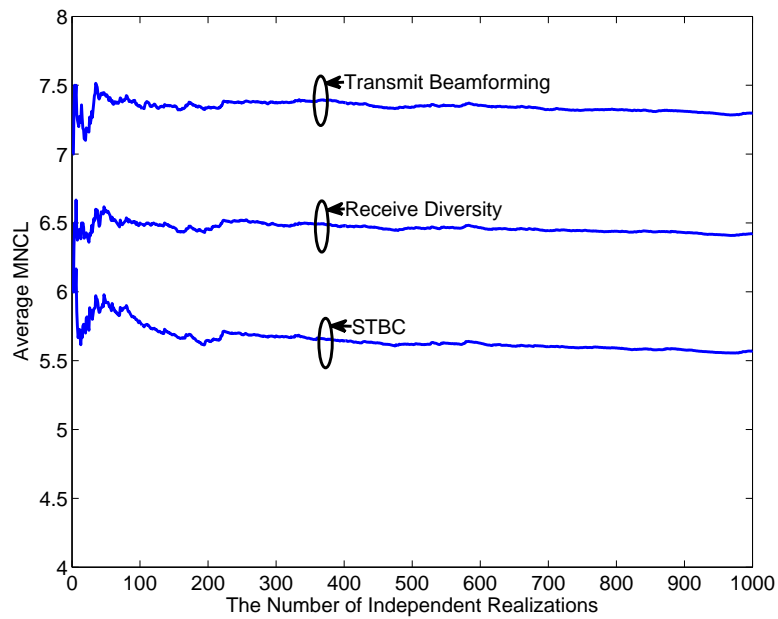}
\caption{Estimate results for the MNCL versus the number of
independent realizations. Assume total 12 transceiver pairs and
$M=4$.} \label{Fig5}
\end{figure}

\begin{figure}
\centering
\includegraphics[width=4in]{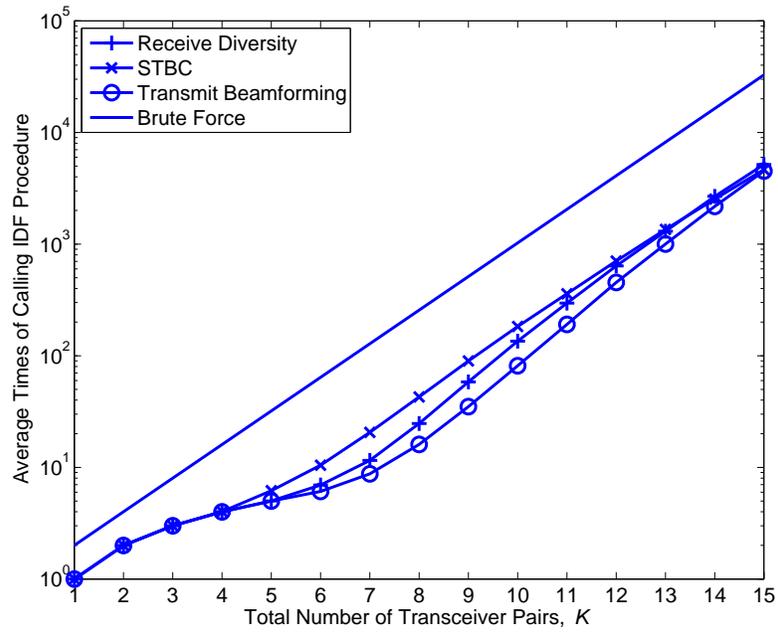}
\caption{Average number of times the IDF procedure was called versus
the number of transceiver pairs for $M=4$.} \label{Fig6}
\end{figure}

\begin{figure}
\centering
\includegraphics[width=4in]{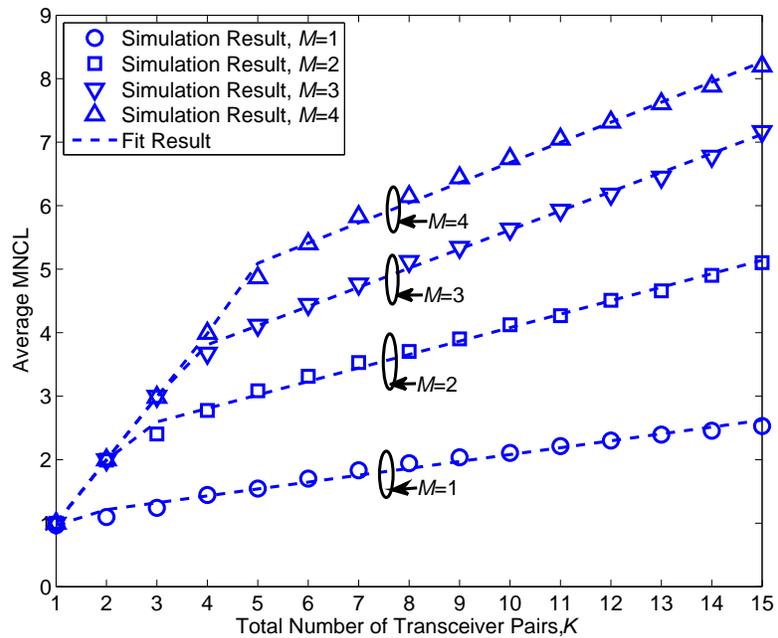}
\caption{Simulation and approximation results for average MNCL with
Transmit Beamforming. Assume 12 pairs equipped with 4 receive
antennas.} \label{Fig7}
\end{figure}

\begin{figure}
\centering
\includegraphics[width=4in]{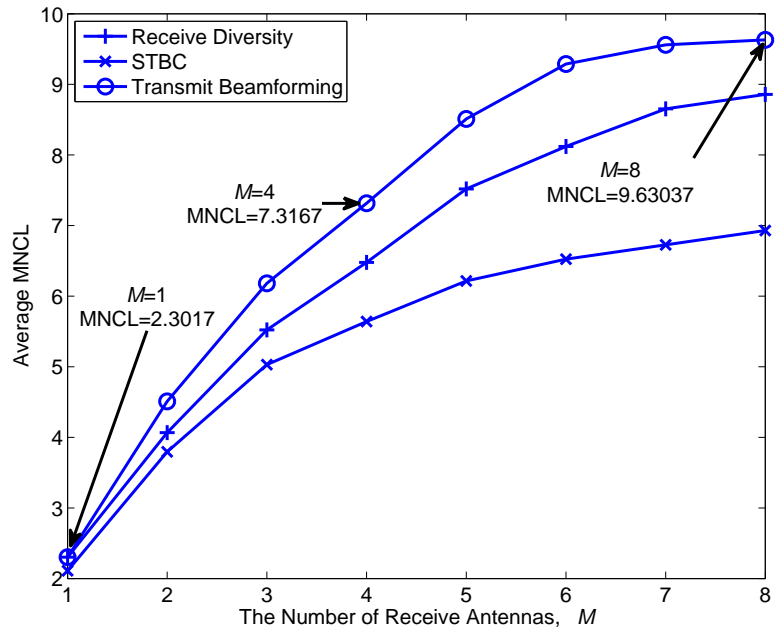}
\caption{Results under different numbers of receive antennas. Assume
total 12 pairs. Up to 8 receive antennas are simulated.}
\label{Fig8}
\end{figure}

\begin{figure}
\centering
\includegraphics[width=4in]{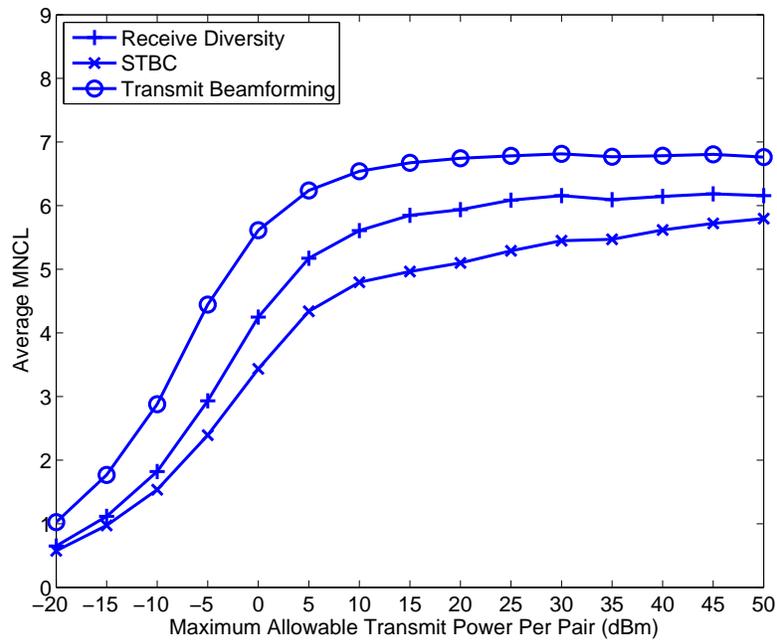}
\caption{Average MNCL versus different maximum allowable transmit
power per pair. Assume a total of 10 pairs, $M$ is fixed as 4.}
\label{Fig9}
\end{figure}

\end{document}